\documentclass{osa-article}

\usepackage[utf8]{inputenc} 
\usepackage{hyperref}       
\usepackage{url}            
\usepackage{amsfonts}       
\usepackage{nicefrac}       
\usepackage{graphicx}
\usepackage{dcolumn}
\usepackage{bm}
\usepackage{upgreek}
\usepackage{amsmath}
\usepackage{lineno}
\usepackage{bm}
\usepackage{graphicx}

\usepackage{algorithmic}

\usepackage{multicol}
\usepackage{multirow}

 \journal{osac}

\articletype{Research Article}

\begin{document}

\title{Efficient Reservoir Computing using Field Programmable Gate Array and Electro-optic Modulation}

\author{Prajnesh Kumar,\authormark{1} Mingwei Jin,\authormark{1},Ting Bu\authormark{1}, Santosh Kumar\authormark{1} and Yu-Ping Huang\authormark{1,2,*}}

\address{\authormark{1}Physics Department and the Center for Quantum Science and Engineering, Stevens Institute of Technology, Hoboken, NJ, 07030, USA.\\
\authormark{2}Department of Electrical and Computer Engineering, Stevens Institute of Technology, Hoboken, NJ, 07030, USA}

\email{\authormark{*}yhuang5@stevens.edu} 



\begin{abstract}
We experimentally demonstrate a hybrid reservoir computing system consisting of an electro-optic modulator and field programmable gate array (FPGA). It implements delay lines and filters digitally for flexible dynamics and high connectivity, while supporting a large number of reservoir nodes. To evaluate the system's performance and versatility, three benchmark tests are performed. The first is the $\bm{10^{th}}$ order Nonlinear Auto-Regressive Moving Average test (NARMA-10), where the predictions of 1000 and 25,000 steps yield impressively low normalized root mean square errors (NRMSE's) of 0.142 and 0.148, respectively. Such accurate predictions over into the far future speak to its capability of large sample size processing, as enabled by the present hybrid design. The second is the Santa Fe laser data prediction, where a normalized mean square error (NMSE) of 6.73$\bm{\times10^{-3}}$ is demonstrated. The third is the isolate spoken digit recognition, with a word error rate close to 0.34\%. Accurate, versatile, flexibly reconfigurable, and capable of long-term prediction, this reservoir computing system could find a wealth of impactful applications in real-time information processing, weather forecasting, and financial analysis.
\end{abstract}


\section{Introduction}

Modern computers based on von Neumann architectures have been designed for digital information
processing and generic computational tasks there upon. With more and more problems being identified as computationally complex and/or excessively time consuming, alternative computational paradigms are actively explored to solve these problems as they increasingly emerge
\cite{liu_dna_2000,McMahon2016614,7549034, AMATO201347,Kumar2020}. To that end, a promising approach is to develop brain-inspired architectures for information processing, including a variety of neural networks \cite{article3,LeCun2015,Mnih2015,Silver2016}. Yet, those architectures are still implemented through computer simulations on Von Neumann computers, thus fundamentally subject to the latter's limitations in speed, parallelism, etc.

Recently, there has been increasing interest on artificial neural networks using optics, leveraging its remarkable speed, multiplexing capability, and little heat deposition \cite{Coherent,Spiking}. 
Usually, these optical neural networks are wholly  trained with a known data set to optimize 
their connectivity and parameters through nonlinear layers \cite{9064516,Bu_2020}. However, such training is usually energy and time consuming, and its efficiency varies by the complexity of task, the size of the network, the nonlinearity and connectivity between the nodes. Also, it is conducted offline using digital computers, thus failing to account for the uncertainties and fluctuations that are inevitable in any optics systems, especially when the optical networks are complex and involve many free-space parts \cite{Zuo:19,Zhou:20}. 

In an effort to address the above difficulties, the idea of reservoir computing (RC) was proposed and widely explored \cite{reservoir_ini, Appeltant2011, antonik_brain-inspired_2017,brunner_tutorial_2018,Chaos_review}. RC origins from liquid-state machine (LSM) \cite{LSM} and echo state networks (ESN) \cite{Jaeger78}. Its realizations are generally composed of three parts: an input layer, a reservoir layer, and an output layer. The input signals are first fed into the input layer, then mapped to the reservoir layer, which contains a large number of interconnected nonlinear nodes and performs nonlinear transformation. Afterwards, response is readout with linear weighted sum of the reservoir states in the output layer. Unlike other kinds of recurrent neural networks that are notoriously hard to be trained, here only the output weights are needed to be optimized, making the training much simpler. It is this advantage and its success on lots of time dependent tasks, such as chaotic time series prediction and speech recognition \cite{vandoorne_toward_2008,vandoorne_parallel_2011,duport_all-optical_2012, brunner_parallel_2013,mesaritakis_micro_2013,dejonckheere_all-optical_2014,arik_wave-based_2015,mesaritakis_high-speed_2015,larger_high-speed_2017,liu_waveform_2017,van_der_sande_advances_2017,hou_prediction_2018,lugnan_photonic_2020}, that draw much attention across different application areas.  


The first RC system was implemented using analog electronics with a single nonlinear node and delayed feedback \cite{Appeltant2011}. Since then, digital electronics are increasingly adopted 
\cite{Daniel2018,Bogdan2018,Fischer_2019}. Among them, field programmable gate array (FPGA) electronics can make the system compact, stable, low-cost and easily configurable with various commercial systems for real-time information processing. More recently, hybrid opto-electronic feedback systems with FPGA's have reformulated the development of a coherent Ising machine for solving computationally hard optimization problems \cite{Bohm2019} and to build a large networks of identical nodes with arbitrary topology for cluster synchronization, chimera states \cite{Joseph2017} and laminar chaos \cite{Rajarshi_PRL}. 


Here, we add to the exciting progress made in the RC field and experimentally demonstrate a fully-packaged opto-electronic RC system consisting of a Mach-Zehnder electro-optic modulator (EOM) and a FPGA circuit. In our design, both the delay line and filters are implemented digitally within FPGA, which renders the whole system compact and immune to optical drifts and noise, especially compared to fiber-optical realizations \cite{larger_photonic_2012,Duport-fully-analogue-2016,bao_efficient_2019}. Leveraging the filters inside the FPGA, we are able to achieve more dynamics and connections between reservoir nodes. 
To characterize the system, we run three benchmark tasks: the $10^{th}$ order Non-Linear Auto-Regressive Moving Average test (NARMA-10), the Santa Fe laser data prediction, and the isolate spoken digit recognition. All exhibit exceptionally high performance, which indicates the robustness and versatility of this RC system. 

The remainder of the paper is organized as follows. Section \ref{Section2} discusses the basics theory of opto-electronic RC system using delay feedback. In Section \ref{Section3}, we show the experimental setup which covers detailed block level FPGA implementations. Subsequently, Section \ref{Section4} shows the experimental results to discuss three benchmark tests for evaluating the performance of RC system and finally, Section \ref{Section5} concludes this paper.

\section{Theory}
\label{Section2}
\begin{figure}[htbp]
    \centering
    \includegraphics[width=8.5cm]{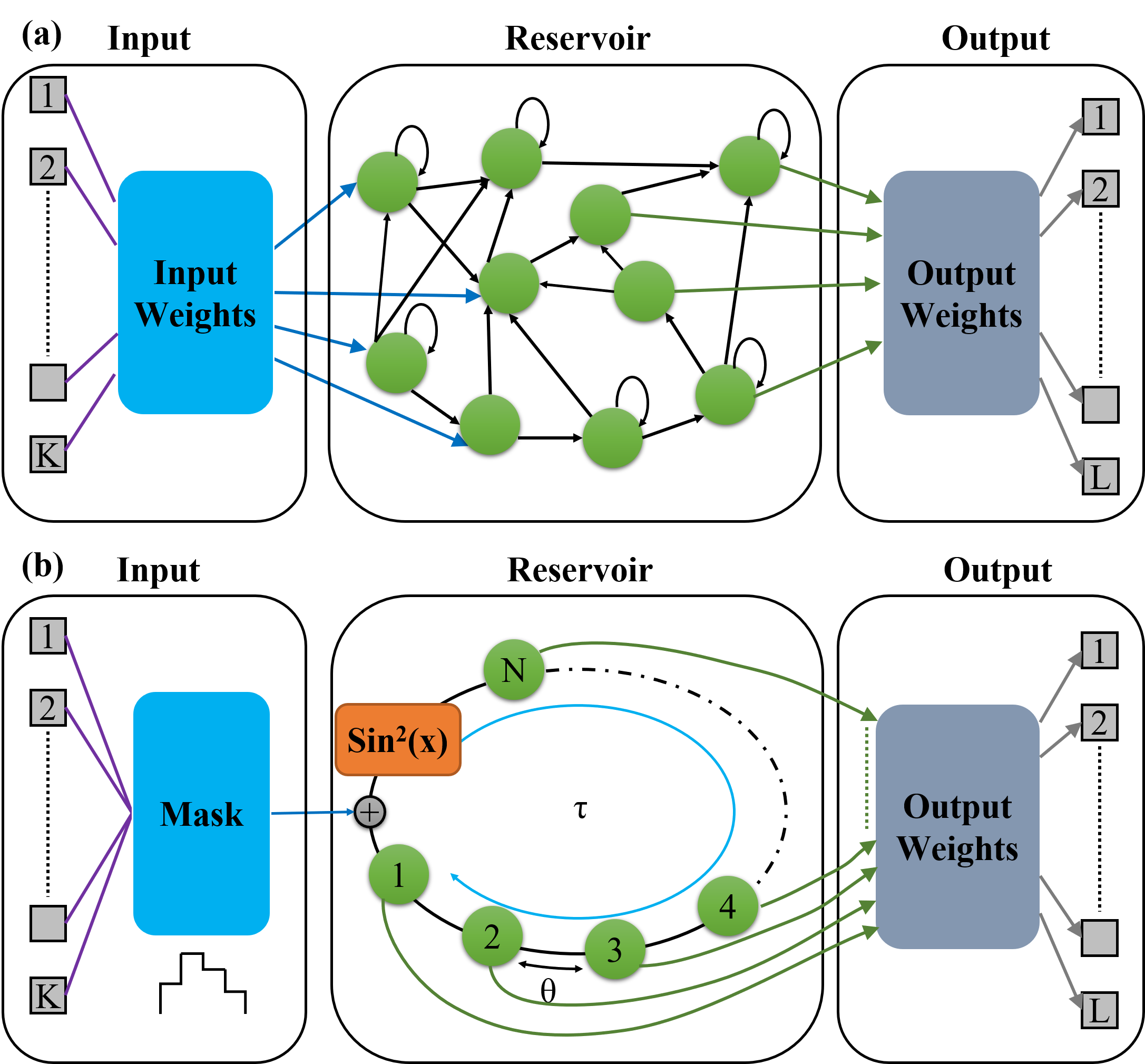}
    \caption{(a) Conventional reservoir computing (RC) model and (b) time-delay based RC model.}
    \label{RC_block}
\end{figure}

Figure \ref{RC_block}(a) shows the conventional RC model which consists of an input, a reservoir, and an output layer. The input layer consists of input vector \textbf{u}($t$) of length $K$ which are fed into the reservoir via fixed but random input weighted connection. These weights will scale the input differently for different reservoir nodes.
The reservoir layer contains a large number of recurrent and randomly interconnected nonlinear nodes. Reservoir layer non-linearly projects the inputs to high-dimensional state space. The dynamics of the reservoir also exhibit a fading memory where the current reservoir state is influenced only by the recent past. The dynamics of states in the reservoir layer is given by 
\begin{equation}\label{eq_re}
x_i(t)=f_{NL}\Big(\sum_k w_{ik} x_k(t-\tau)+\sum_j M_{ij} u_j(t)\Big),
\end{equation}
where $f_{NL}$ is the non-linear function, $x_i(t)$ is the reservoir state of $i^{th}$ node at time $t$, $w_{ik}$ is the nodes inter-connection matrix, $M_{ij}$ is the input weight matrix and $u_j(t)$ is the $j^{th}$ input.

At the output layer, the response is readout by linear weighted sum of node states which described as 
\begin{equation}\label{eq_out}
\hat{y}_j(t)=\sum_iW_{ij} x_i(t),
\end{equation}
where $W_{ij}$ being the output weights. 
The output weights $W_{ij}$ are updated after training to make the outputs $\hat{y}_j(t)$ as close as possible to the target values $y_j(t)$ using optimization methods such as linear regression, ridge regression etc.
Above RC architecture requires very large number of interconnected physical nodes. In contrast, the time-delay based RC uses virtual nodes that are temporally spaced and has only one nonlinear node as shown in Fig.~\ref{RC_block}(b). The non-linearity is implemented electro-optically, using an EOM. The input is preprocessed before injecting into nonlinear node, this procedure is  referred as masking. The preprocessed input signal is time-multiplexed and injected serially into reservoir. The input vector after preprocessing will resides in total delay time of $\tau$. The temporal spacing between N nodes is given by $\theta=\tau/N$ as shown in Fig. \ref{RC_block}(b). 



\begin{figure}[htbp]
    \centering
    \includegraphics[width=9cm]{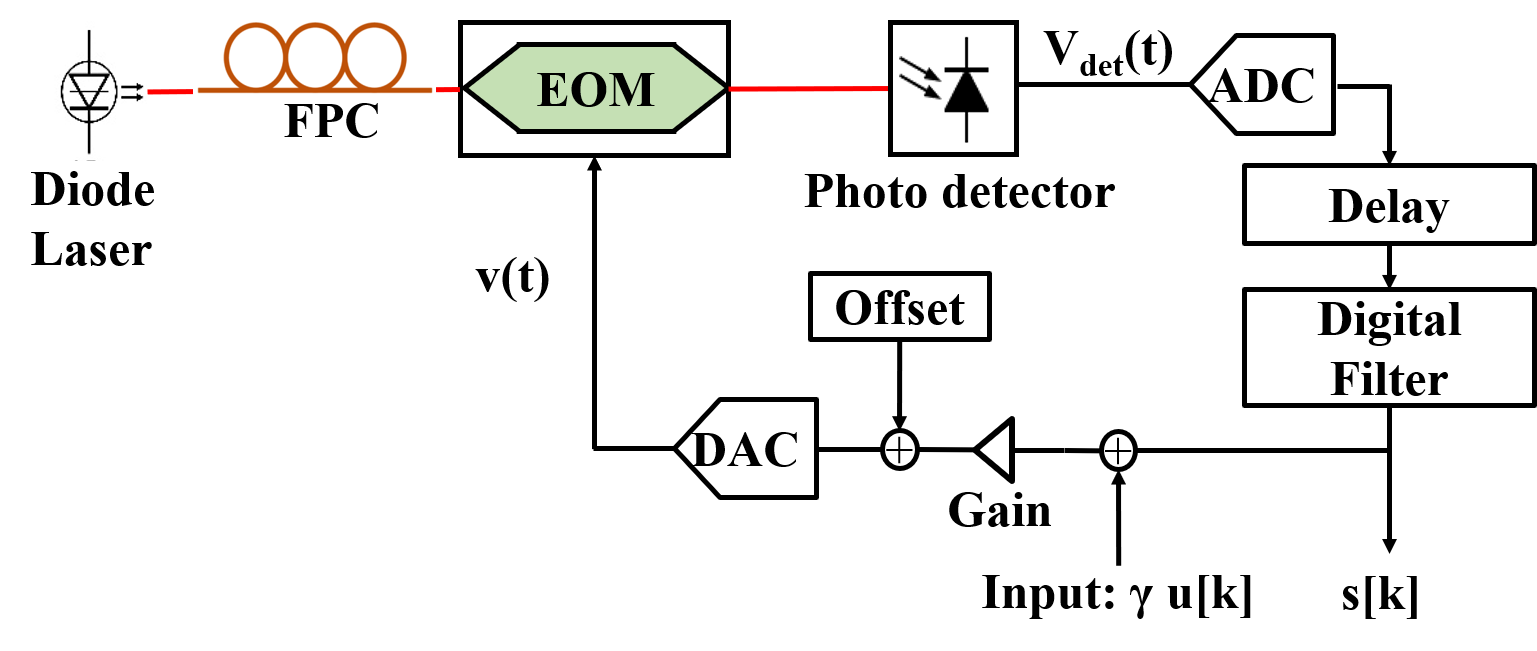}
    \caption{Schematic representation of digital opto-electronic RC}
    \label{AnalogRC}
\end{figure}
Figure~\ref{AnalogRC} shows the schematic of our digital opto-electronic RC implementation of delay feedback system. 
The optical power at the output of EOM is given by a transfer function
\begin{equation}
P(t) \sim \sin^2[\pi v(t)/(2V_\pi)+\phi], 
\label{mzi_transfer_func}
\end{equation}
where $v(t)$ is the RF signal applied to EOM, $V_\pi$ is the 
$\pi$- shift voltage and $\phi$ is the bias offset. 

The photo-detector generates voltage in response to 
the applied laser power which is given by 
\begin{equation}
V_{det}(t) = \alpha \eta R_t P(t), 
\label{photo_detector_response}
\end{equation}
where $\alpha$ is the total insertion loss of the EOM, $\eta$ is the responsivity of the detector and $R_t$ is transimpedance gain of the amplifier. The RF input to EOM is 
\begin{equation}
v(t)=G[s(t) + \gamma u(t)],
\label{RF_EOM}
\end{equation}
where $G$ is the overall forward gain of the system,$\gamma$ is the input scaling factor and $u(t)$ is the input to the reservoir. $s(t)$ is the output of the filter after delay, which is convolution of the delayed detector output and filter impulse response function $h(t)$, as
\begin{equation}\label{eqFilter_ouput}
\begin{split}
s(t)&= h(t)*V_{det}(t-\tau),  \\
\end{split}
\end{equation}
with $\tau$ the total delay.

Taking all into account, the equation of the states reads
\begin{equation}\label{eq022}
\begin{split}
x(t)&=\beta h(t)*\sin^2 \left[x(t-\tau) + \gamma' u(t-\tau)+\phi \right]  \\
\end{split}
\end{equation}
with $x(t)=\pi G s(t)/2V_{\pi}$ , $\beta=\pi G \eta R_t\alpha |E_0|^2/2V_{\pi}$, and $\gamma '=G\gamma/2V_\pi$.
It is then discretized as 
\begin{equation}\label{eq023}
\begin{split}
x[k]        &=\beta h[k]*\sin^2 \left(x[k-N] + \gamma' u[k-N]+\phi \right),  \\
\end{split}
\end{equation}
where $k=t/dt$ is discrete sample obtained with sampling period $dt$ and $\tau$ is integer multiple $N$ of $dt$. For digital filter, $h[k]$ has a finite length of $M$ with $M\leq N$. Thus, Equation ~(\ref{eq023}) is in an explicit form of 
\begin{equation}\label{eq024}
\begin{split}
x[k]=\beta  \sum_{j=0}^{M-1} h[j]\sin^2 \left(x[k-N-j] + \gamma' u[k-N-j]+\phi \right).  
\end{split}
\end{equation}




From Eq. (\ref{eq024}), we can see that the states are coupled using filter coefficients. Digital filter gives the flexibility to implement different network topology \cite{larger_photonic_2012}.


\section{Experimental Setup}
\label{Section3}
\begin{figure}[htbp]
    \centering
    \includegraphics[width=8cm]{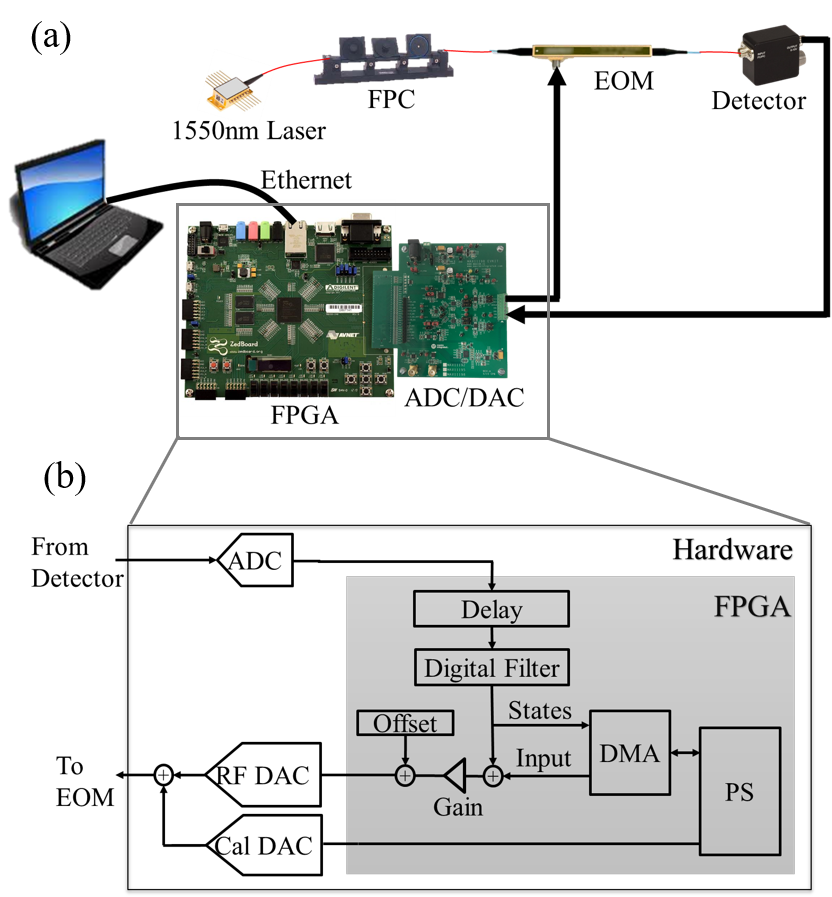}
    \caption{ (a) Experimental setup for the present opto-electronic RC and (b) the block diagram of the hardware implementation using FPGA board, and ADC/DAC. }
    \label{BlockDiagram}
\end{figure}

The block diagram of our experiment setup is depicted in Figure \ref{BlockDiagram}(a). A fiber-coupled laser diode (JDS Uniphase CQF975/58) provides a continuous wave beam at $\sim$1550 nm, which is passed through a fiber polarization controller (FPC) before coupling to an EOM (Lucent Technologies, 2623NA). 
The EOM takes RF input signal from the arbitrary waveform generated using RF Digital to Analog converter (DAC) and modulates the laser beam intensity. 
The modulated intensity is converted to electrical signal using a photodetector (Thorlabs, DET10C2), whose output is digitized using an analog to digital converter (ADC) with 1 Msps sampling rate controlled by FPGA. A Personal computer (PC) provides the preprocessed data via Ethernet interface for this RC system.
Figure \ref{BlockDiagram}(b) shows the FPGA blocks implemented on a Zedboard (Zynq-7000) development platform.
The RC logic is implemented using Verilog programming language.
The FPGA interfaces to ADC (Maxim Integrated MAX11198, 16bits) and RF DAC (Analog Devices D5541A,16bits) using Serial Peripheral Interface.
The PC preprocess the input signal using mask and input scaling factor, which is then sent to a Zedboard via Ethernet interface. Processing System (PS) in Zedboard shares data with Direct Memory Access (DMA) to stream the input to the FPGA logic. The feedback signal from filter block and the streamed input are added and passed through programmable gain and offset block before converting to electrical signal using DAC. Here, the gain and offset are key parameters for tuning RC system performance. Similarly streaming data from ADC is fed into programmable delay block with delay limited to 1000 units, each delay representing 1 virtual node spacing. The delayed signal is then passed to finite digital bandpass filter which has 400 filter taps. The filtered data implements the Eq. (\ref{eq024}) which has state information. PS will then transmit this data to PC via Ethernet interface. Calibration DAC (Maxim Integrated MAX5316, 16bits) is used  to compensate for the bias drift of EOM.

\section{Results}
\label{Section4}
\subsection {NARMA10}

NARMA is one of the most widely used benchmarks in RC\cite{Jaeger2003}. NARMA10 is a discrete time nonlinear task with $10^{th}$ order lag. Series $y_{k+1}$ is generated through a recursive formula
\begin{equation}\label{NARMA10}
y_{k+1} =0.3y_k+0.05y_k \sum_{i=0}^{9}y_{k-i} + 1.5u_k u_{k-9}+0.1.  
\end{equation}
The input $u_k$ is drawn from a uniform distribution in the interval $[0,0.5]$. Due to non-linearity and long time lag, NARMA10 poses a challenge for any computation system.

To characterize the performance of the RC, a normalized root mean squared error (NRMSE) between the target and predicted value is calculated as
\begin{equation}\label{NRMSE}
\textrm{NRMSE} =\sqrt{\frac{1}{m}\frac{\sum_{k=0}^{m}(\hat{y}_{k}-y_{k})^2}{\sigma^2(y_{k})}},
\end{equation}
where $y_{k}$ is the target, $\hat{y}_k$ is the prediction, $m$ is the total number of samples in the target and $\sigma$ denotes the standard deviation of the target. 

\begin{figure}[htbp]
    \centering
    \includegraphics[width=9cm]{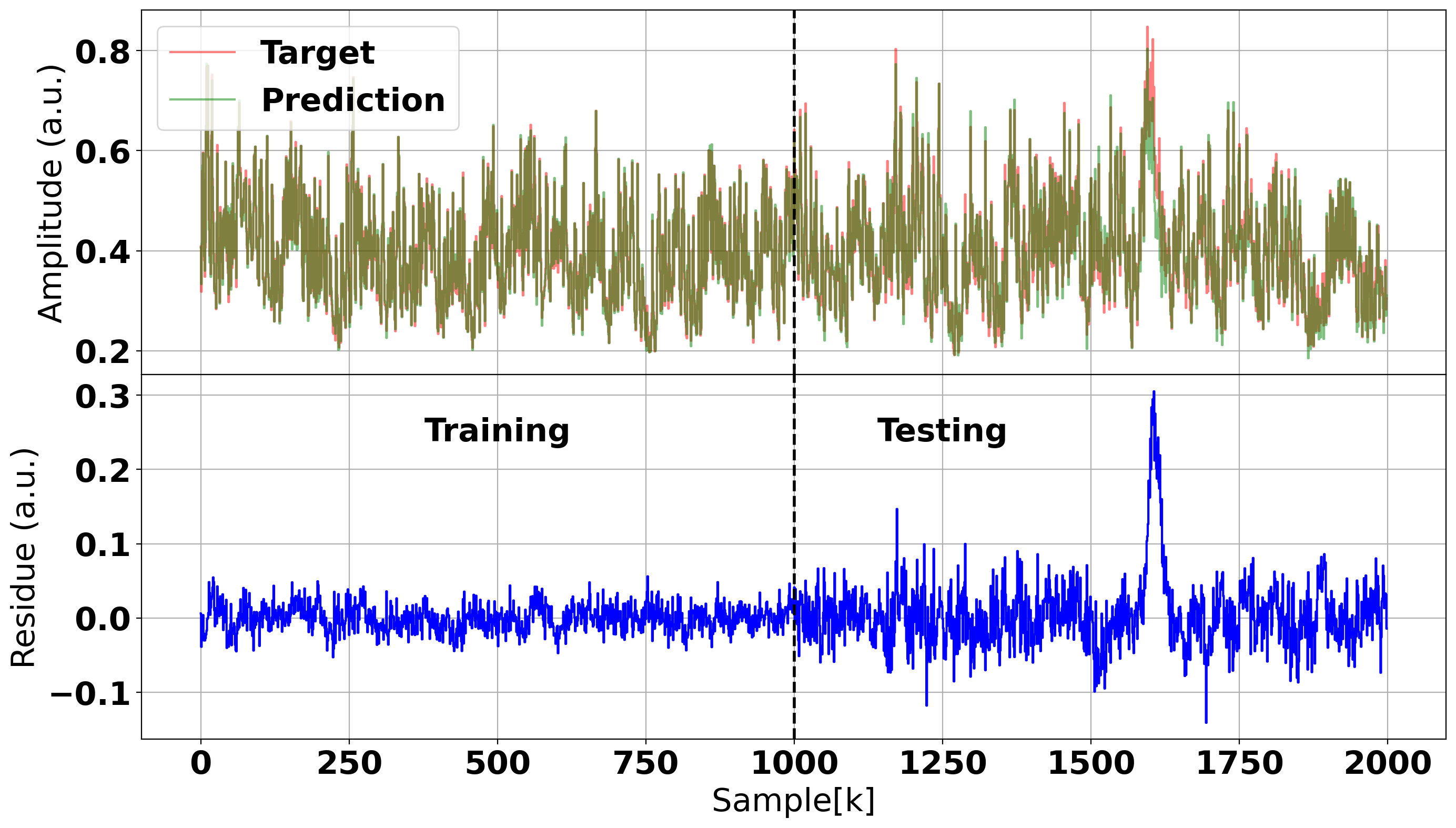}
    \caption{NARMA10 benchmark results: (a) shows the amplitude of the target (red line) and the predicted signal (green line) vs index of 2000 samples and (b) shows the normalized residue of target and predicted signal.}
    \label{Narma10BenchmarkResults}
\end{figure}

By sweeping the system parameters, the optimum operating point is identified at  gain $G =0.58$,  input scaling $\gamma =0.5$, bias $\phi=0.1\pi$, total of $N=400$ virtual nodes. The results of benchmark are tabulated in Table \ref{NARMA10Table}. Figure \ref{Narma10BenchmarkResults} plots the target vs prediction for 2000 samples. Also plotted is the residue $\mathrm{\bf R}$, defined as the difference between target and estimate normalized by the mean of target 
\begin{equation}\label{residue}
\mathrm{\bf R}[k]=m(y_{k}-\hat{y}_{k})/\sum_k y(k).
\end{equation}
As seen, the present RC performs remarkably well over long term prediction. For 25000 samples with training size of 5200, the mean NRMSE is 0.148.
\begin{table}[htbp]
\caption{\bf NARMA10 benchmark results with 400 nodes. }
\label{NARMA10Table}
\centering
\begin{tabular}{cccc}
\hline
\textbf{Samples} & \textbf{Mean} & \textbf{Std} & \textbf{Min} \\ 
\hline
$2000$              &$0.142$         &$6.47\times 10^{-3}$     &$0.129$        \\ 
$25000$             &$0.148$         &$5.90\times 10^{-3}$     &$0.139$        \\ 
\hline
\end{tabular}

\end{table}

\subsection{Santa Fe laser data prediction}
The data set A from the 1994 time series prediction competition organized by the
Santa Fe Institute was an time series obtained from measuring a NH3 laser and it is good example of realistic data \cite{RePEc:eee:intfor:v:10:y:1994:i:1:p:161-163}. Santa Fe laser data prediction is one step series prediction and we use 4000 data points in our test case. The performance is measured using NMSE. 

\begin{figure}[htbp]
    \centering
    \includegraphics[width=9cm]{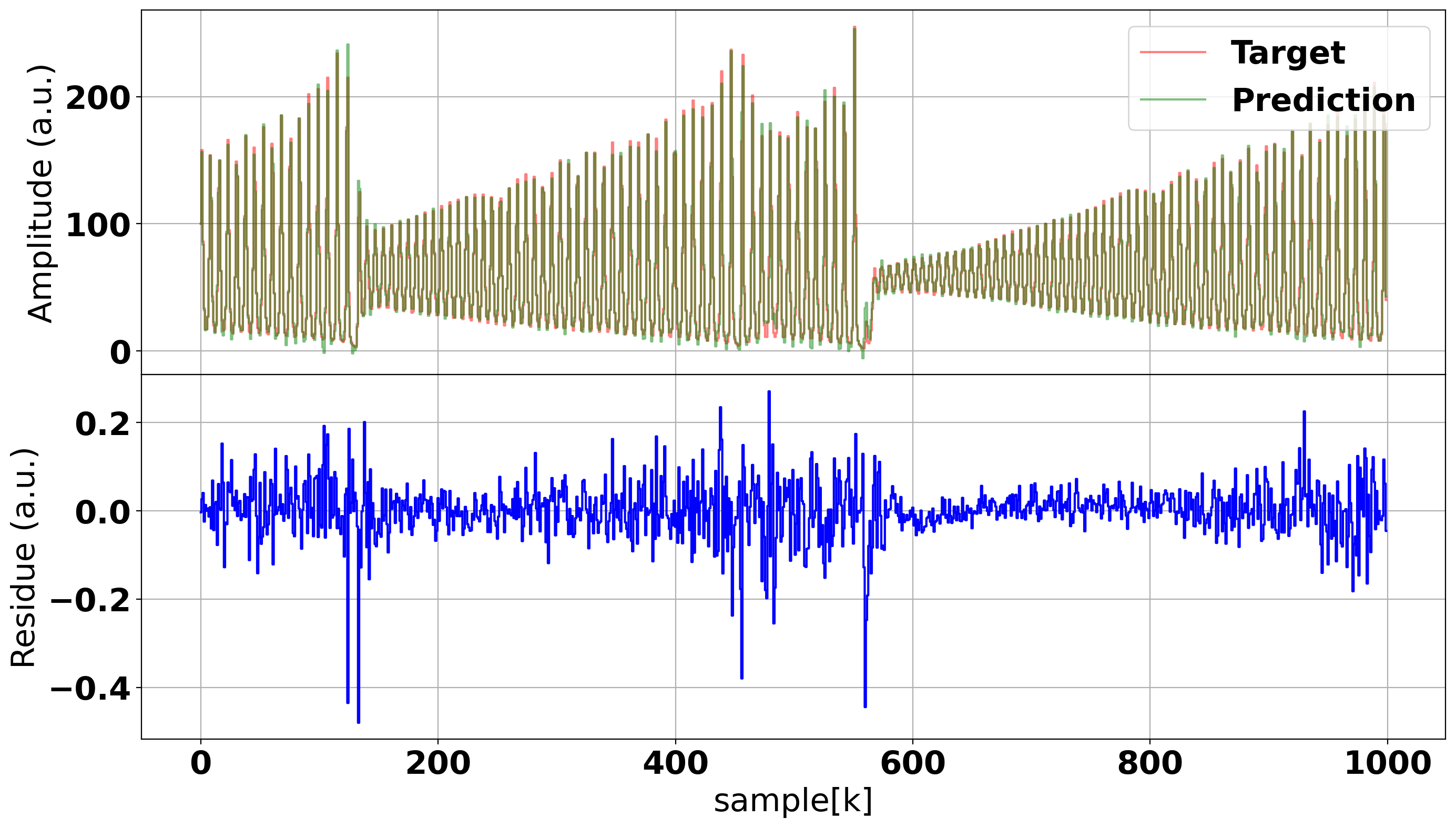}
    \caption{Santa Fe benchmark results: (a) shows the amplitude of the target (red line) and the predicted signal (green line) as a function of sample size and (b) shows the normalized residue of target and prediction.}
    \label{SantaFeOneStepBenchmarkResults}
\end{figure}

By sweeping the system parameters, the optimum operating point is identified at  gain $G =0.58$,  input scaling $\gamma =0.0029$, bias $\phi=0.1\pi$, total of $N=400$ virtual nodes for a data points of 4000. Results of the experiment is listed in Table \ref{SanaFeSingleStepResultTab}. Figure \ref{SantaFeOneStepBenchmarkResults} shows plots of the target and prediction, along with the Residue for the first 1000 samples. 


\begin{table}[htbp]
\caption{\bf Santa Fe laser data single step benchmark results. }
\label{SanaFeSingleStepResultTab}
\centering
\begin{tabular}{cccc}
\hline
\textbf{Nodes\#} & \textbf{Mean} & \textbf{Std} & \textbf{Min} \\ 
\hline
$400$                   &$2.71 \times 10^{-2}$      &$6.20\times 10^{-4}$     &$2.54\times 10^{-2}$     \\ 
$950$                   &$6.73\times 10^{-3}$      &$3.34\times 10^{-4}$     &$5.66\times 10^{-3}$     \\ 
\hline
\end{tabular}
\end{table}

\subsection {Isolate spoken digit recognition}
Speech recognition is a commonly used benchmark for testing the performance of neural networks. Acoustic feature are more pronounced in frequency domain compared to time domain. Hence the input audio data are first pre-processed by decomposing the time-domain information into frequency-time information. In our implementation we are using Lyons cochlear ear model to get the cochleagram which mimics filtering that occurs in nature \cite{1171644}. 

\begin{figure}[htbp]
    \centering
    \includegraphics[width=9cm]{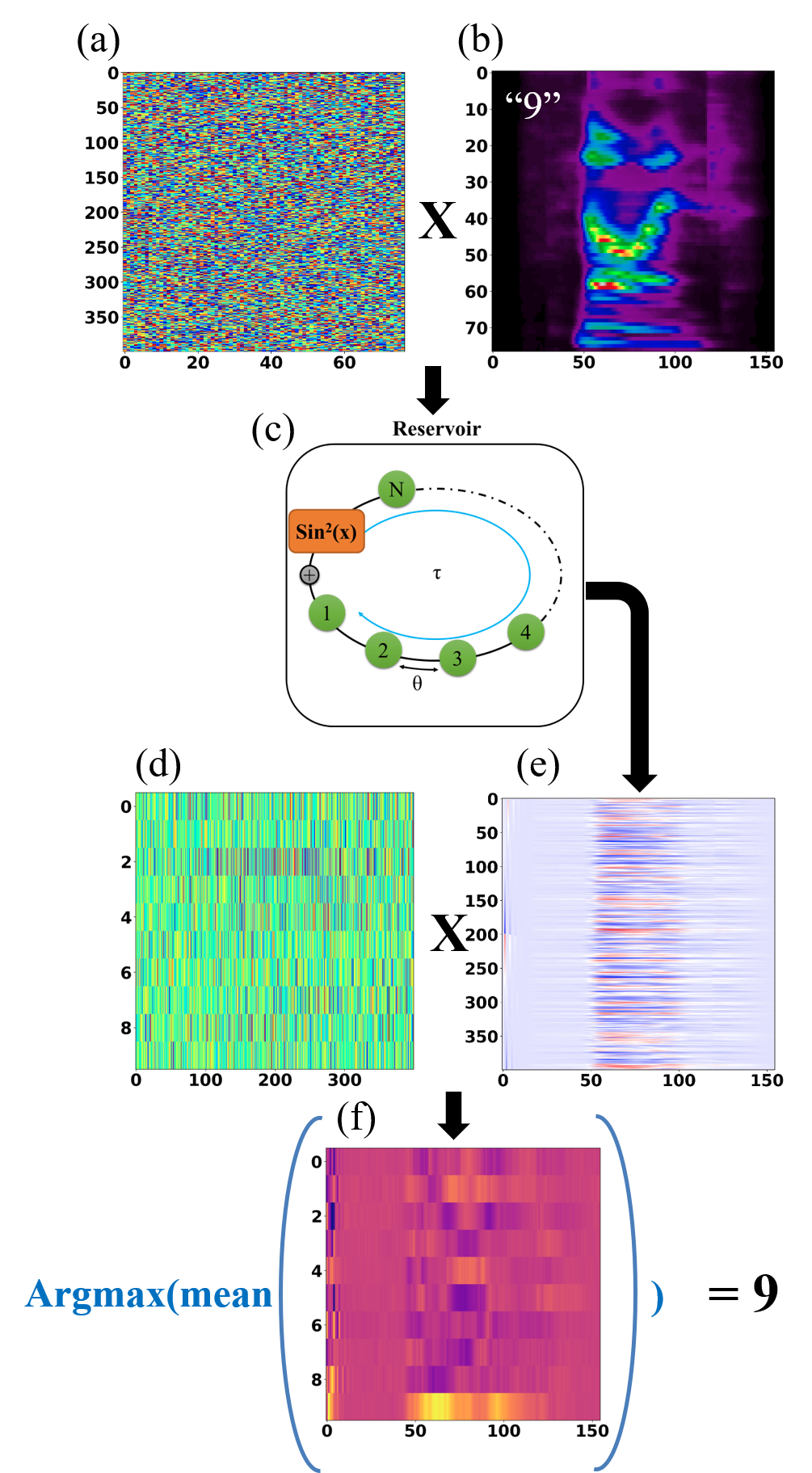}
    \caption{Graphical illustration of isolated spoken digit recognition task. (a) Uniformly distributed input mask with values in range [-1,1] and dimension $N$x$N_{ch}$ where $N=400$ Nodes and $N_{ch}=77$ (b) Cochleagram generated from audio file of dimension $N_{ch}$x$N_T$ corresponding to digit 9. (c) Resultant product of input mask and cochleagram is serialized and injected into reservoir. (d) Output layer weights of dimension $N_{d}$x$N$ where number of output classes is $N_{d}=10$  (e) The output from reservoir is serially captured and reshaped to $N$x$N_T$, this will give node matrix. (f) Product of output weight with reservoir state matrix give estimation matrix. Output class is predicted by getting the maximum argument corresponding to row-wise mean of estimation matrix.
    }
    \label{isolatedSpokenDigitProcess}
\end{figure}

The input audio data for this experiment is from AudioMNIST database \cite{becker2018interpreting}. This is a free and open database that contain 30,000 spoken digits(0-9) audio samples of 60 different speakers with a sampling rate of 48 kHz. To use these audio files with our RC, we first down-sample the audio to 12 kHz using the librosa package \cite{brian_mcfee_2020_3955228} and pad audio files with random length zero at beginning and end to make total sample size of $12,000$. Next we generate the cochleagram as shown in Fig. \ref{isolatedSpokenDigitProcess}(b), which is calculated using ``lyon 1.0.0" python package. In this test, we consider first 5 speakers from the AudioMNIST database to create 20 balanced subset, each containing 5 speaker $\times$ 10 digit $\times$ 2 utterance each, i.e. 100 audio sample. The training and testing dataset will contains 5 such subset with the first 4 used for training and the last one for testing. In-order to obtain unbiased result, 50 datasets are formed using random combinations of 5 subsets from 20 subsets and the word error rate (WER) is measured for each combinations. 

Figure \ref{isolatedSpokenDigitProcess} illustrate the flow for spoken digit recognition task. The input mask is a real valued matrix of dimension $N \times N_{ch}$ with uniform distribution in [-1,1], where $N$ is number of virtual nodes and $N_{ch}$ is the number of cochleagram channels. Figure \ref{isolatedSpokenDigitProcess}(b) shows the cochleagram for digit 9 with dimension $N_{ch} \times N_T$ where $N_T$ is the index of new time representation that depends on decimation factor. As shown in Fig. \ref{isolatedSpokenDigitProcess}(c), the resultant product of input mask and cochleagram is serially injected into reservoir by flattening the matrix. The output of the reservoir is de-serialized to get the reservoir state matrix of dimensions $N \times N_T$, as shown in Fig.~\ref{isolatedSpokenDigitProcess}(e).
The estimation matrix is calculated by multiplying the output weight matrix with reservoir state matrix. The Output class is predicted by getting the maximum argument corresponding to row-wise mean of estimation matrix \cite{AbreuAraujo2020}.

After tuning the system parameters and based on their error evaluations, 
we found the optimum operating point at gain $G =0.5$,  input scaling $\gamma =300$, bias $\phi=0.35$, total of $N=400$ virtual nodes. Results of experiment are listed in table \ref{IsolatedspokendigitrecogntionTab}. 

\begin{table}[htbp]
\caption{\bf WER for Isolated spoken digit recognition task for training and testing phase.}
\label{IsolatedspokendigitrecogntionTab}
\centering
\begin{tabular}{ccc}
\hline
& \textbf{Training WER} & \textbf{Testing WER} \\ 
\hline
\textbf{Mean} & $0.05\%$                 & $0.34\%$                \\ 
\textbf{Std}  & $0.13\%$                 & $0.51\%$\\ 
\hline
\end{tabular}

\end{table}

\begin{table}
\centering
\caption{\bf Performance Metric Comparison of Various RC Systems.   AWG:Arbitrary Wavefrom generator, DAQ: Data Acquisition.}
\label{compareresults}
\resizebox{\textwidth}{!}{\begin{tabular}{llcll}
\hline
\textbf{Reference} 
& \textbf{Opto-electronic RC system} 
& \textbf{Nodes \#} 
& \textbf{Test case} 
& \textbf{Performance}     
\\ \hline
\textbf{\cite{larger_high-speed_2017}}
& {\begin{tabular}[t]{@{}l@{}}Band-pass, optical phase dynamical
                                    \\ variable, 
                                    \\AWG @ 24 GS/s. 
                                    \\DAQ @ 80 GS/s\end{tabular}}
&{1113}
& \begin{tabular}[t]{@{}l@{}}\textbf{Spoken digit recognition}
                                    \\Frequency Channel =86
                                    \\Training/Testing Size=475/25  digits,
                                    \\repeated 20 times.\end{tabular}
& {WER=$0.04\%$}   \\

\multirow{2}{*}{\textbf{\cite{larger_photonic_2012}}} 
& \multirow{2}{*}{{\begin{tabular}[t]{@{}l@{}}Low-pass, 
                \\voltage dynamic variable\end{tabular}}}
& {400}
& \begin{tabular}[t]{@{}l@{}}\textbf{Santa Fe laser time-series}
\\One step prediction\end{tabular}
& {NMSE=$0.124 \pm 4\times10^{-4}$}       \\

&
& {400}
& \begin{tabular}[t]{@{}l@{}}\textbf{Spoken digit recognition}
                                    \\Frequency Channel=86
                                    \\Training/Testing Size=475/25 digits,
                                    \\repeated 20 times\end{tabular}
& {WER=$0.04\% \pm0.017\%$}      \\

\textbf{\cite{Duport-fully-analogue-2016}}
& \begin{tabular}[t]{@{}l@{}}Band-pass, voltage dynamic variable,
                            \\fully analog system,
                            \\AWG @ 200 MS/s, 16 bits resolution, 
                            \\DAQ @ 200 MS/s, 12 bits resolution.\end{tabular}
& {47}
& \begin{tabular}[t]{@{}l@{}}\textbf{NARMA10}
                                    \\Training/Testing size=1000/1000.\end{tabular}
& {NMSE=$0.230 \pm 0.023$}    \\

\textbf{\cite{Duport-2016}}
& \begin{tabular}[t]{@{}l@{}}Band-pass, voltage dynamic variable,
                            \\Simultaneous task
                            \\AWG @ 200 MS/s, 16 bits resolution.
                            \\DAQ @ 200 MS/s, 12 bits resolution.\end{tabular}
& {50}
& \begin{tabular}[t]{@{}l@{}}\textbf{NARMA10}
                                    \\Training/Testing size=3000/6000.\end{tabular}
& {NMSE=$0.181 \pm 0.013$}\\

\textbf{\cite{Hermans2016}}
& \begin{tabular}[t]{@{}l@{}}Band-pass, voltage dynamic variable, 
                                    \\optimization via backpropagation\end{tabular}
& {80}
& \begin{tabular}[t]{@{}l@{}}\textbf{NARMA10}
                                    \\ Back propogation algorithm with \\20,000 iterations\end{tabular}
& {NRMSE=$0.185$}          \\
\textbf{\cite{soriano_optoelectronic_2013}}
& {Band-pass, voltage dynamic variable}
& {400}
& \begin{tabular}[t]{@{}l@{}}\textbf{Santa Fe laser time-series}
                                \\Training/Testing size=3000/1000.\end{tabular}
& {NMSE=$0.02$}             \\

\textbf{\cite{martinenghi_photonic_2012}}
& {\begin{tabular}[t]{@{}l@{}}Low-pass, 
                                    \\wavelength dynamical variable\end{tabular}}
& {150}
& \begin{tabular}[t]{@{}l@{}}\textbf{Spoken digit recognition}
                                    \\Frequency Channel =86
                                    \\Training/Testing Size=475/25  digits,
                                    \\repeated 20 times.\end{tabular}
& {WER=$0.6\% \pm 0.2\%$}        \\
\multirow{2}{*}{\textbf{\cite{paquot_optoelectronic_2012}}} 
& \multirow{2}{*}{{\begin{tabular}[c]{@{}l@{}}Band-pass, voltage dynamic variable
                                                    \\DAQ @200 MS/s\end{tabular}}}
&{50}
& \begin{tabular}[t]{@{}l@{}}\textbf{NARMA10}
                                    \\Training/Testing size=1000/1000\end{tabular}
& {NMSE=$0.168 \pm 0.015$} \\

&
& {200}
& \begin{tabular}[t]{@{}l@{}}\textbf{Spoken digit recognition}
                                    \\Frequency Channel=86
                                    \\Training/Testing Size=475/25 digits,
                                    \\repeated 5 times.\end{tabular}
& {WER=$0.4\%$}            \\

\multirow{3}{*}{\textbf{This paper}}       
& {\multirow{3}{*}{{\begin{tabular}[t]{@{}l@{}}Band-pass, Voltage dynamic variable,
                                                    \\AWG @ 1 MS/s, 16 bits resolution 
                                                    \\DAQ @ 1 MS/s, 16 bits resolution.\end{tabular}}}} 
& {400}
& \begin{tabular}[t]{@{}l@{}}\textbf{NARMA10}
                                \\Training/Testing size=1000/1000\end{tabular}
& {NRMSE=$0.142 \pm 6.5\times 10^{-3}$}     \\

& \multicolumn{1}{c}{}
& {950}
& \begin{tabular}[c]{@{}l@{}}\textbf{Santa Fe laser time-series}
                                    \\One step prediction 4000 samples\end{tabular}
& {NMSE=$6.73\times 10^{-3}\pm3.3\times 10^{-4}$}  \\
& \multicolumn{1}{c}{}
& {400}
& \begin{tabular}[c]{@{}l@{}}\textbf{Spoken digit recognition}
                                    \\Frequency Channel=77
                                    \\Training/Testing Size=475/25 digits,
                                    \\repeated 20 times\end{tabular}
& {WER=$0.34\%\pm0.51\%$}       
\\ \hline 

\end{tabular}
}

\end{table}

\section{Conclusions}
\label{Section5}
We have experimentally demonstrated an opto-electronic RC system using EOM and FPGA. It takes advantage of electro-optic nonlinear transformation in EOM, and flexible signal generation and controllable timing, highly programmable signal filtering, and stable delay logic implementation in FPGA. The resultant system is a stable and fully functional reservoir computer supporting many nodes, high connectivity, and easily configurable for multifaceted tasks, while allowing online training.  
We have tested it with three benchmark tasks: NARMA-10, SantaFe laser prediction, and Isolate spoken digit recognition. It achieves 0.142 NRMSE for NARMA-10,  6.73$\times10^{-3}$ NMSE for SataFe, and a WER $\sim$ 0.34\% for the speech recognition. These results are compared with the state of the art in Table \ref{compareresults}. As seen, the present RC system tops the prediction accuracy in the first two test, and is the only reported system to perform well in all three tasks. Moreover, it is able to precisely predict 25,000 steps in NARMA-10 series with impressive low NRMSE of 0.148. Such high performance in those various benchmark tests clearly demonstrate the advantages of our system for complex and versatile tasks. 

Our system is currently working at a low speed, which is mostly limited by the sampling rate of ADC and settling time of DAC. A significant speedup is expected by choosing a faster ADC and DAC. Also, it is prospective to replace the existing bulk-optical EOM with photonic integrated chips, where sophisticated and tailored nonlinear transformations can be realized using nested optical circuits on a single chip \cite{Chen,Jin}. Finally, an FPGA-based opto-electronic RC system of this or similar design are easily re-configurable to accommodate even more neurons, high connectivity, and arbitrary topology. Also, the performance and robustness of the present RC could be further improved by online training. Those upgrades will invite important applications in weather and financial forecasting, real-time information processing, and so on.

\section{Disclosures}






\noindent\textbf{Disclosures.} The authors declare no conflicts of interest.





\bibliography{references}


\end{document}